\documentclass[twocolumn,showpacs,amsmath,amssymb,superscriptaddress]{revtex4}

\usepackage{epsfig,psfrag}
\usepackage{dcolumn}
\usepackage{bm}
\usepackage{graphicx}
\usepackage{color}
\newcommand{\beq}{\begin{equation}}
\newcommand{\eeq}{\end{equation}}
\newcommand{\beqr}{\begin{eqnarray}}
\newcommand{\eeqr}{\end{eqnarray}}
\newcommand{\e}{{\epsilon}}

\newcommand{\phib}{{\bar{\phi}}}
\newcommand{\zb}{{\bar{z}}}

\def\tH{{\tilde{H}}}

\def\tw{{\tilde{w}}}

\def\he{{\hat e}}

\def\bn{{\mathbf n}}

\def\bq{{\mathbf q}}

\def\br{{\mathbf r}}

\def\bR{{\mathbf R}}

\def\bk{{\mathbf k}}

\def\lam{\lambda}

\def\half{{1\over2}}

\def\eqa{\begin{eqnarray}}
\def\eea{\end{eqnarray}}

\def\cK{{\cal K}}
\def\cH{{\cal H}}

\def\vphi{{\varphi}}
\def\dotz{{{\dot{z}}}}
\def\dotphi{{\dot{\phi}}}
\def\dotzeta{{\dot{\zeta}}}
\def\dotvphi{{\dot{\varphi}}}

\def\ssc{Sol. State. Commun.}
\def\njp{New Jour. Phys.}

\def\jpc{{Jour. Phys. C}}
\def\npb{{Nucl. Phys. B}}
\begin{document}

\title{Bilayer Quantum Hall Ferromagnet in a Periodic Potential}
\author{Jianmin Sun}
\affiliation{Department of Physics, Indiana University, Bloomington, IN}
\author{Ganpathy Murthy}
\affiliation{Department of Physics and Astronomy,
University of Kentucky, Lexington KY 40506-0055}
\author{H.A. Fertig}
\affiliation{Department of Physics, Indiana University, Bloomington, IN}
\author{Noah Bray-Ali}
\affiliation{Department of Physics and Astronomy,
University of Kentucky, Lexington KY 40506-0055}

\date{\today}
\begin{abstract}
The bilayer quantum Hall system at a total filling of $\nu_T=1$ has
long resisted explanation in terms of a true counterflow superfluid,
though many experimental features can be seen to be ``almost'' that of
a superfluid. It is widely believed that quenched disorder is the root
cause of this puzzle. Here we model the nonperturbative effects of
disorder by investigating the $\nu=1$ bilayer in a strong periodic
potential. Our model assumes that fermions are gapped and real spins
are fully polarized, and concentrates on the pseudospin variable (the
layer index), with the external potential coupling to the topological
(Pontryagin) density of the pseudospin. We find that as the potential
strength increases, there are ground state transitions in which the
topological content of the pseudospin configuration changes. These
transitions are generically weakly first-order, with a new
quadratically dispersing mode (in addition to the linearly dispersing
Goldstone mode) sometimes becoming nearly gapless near the
transition. We show that this leads to strong suppressions of both the
Kosterlitz-Thouless transition temperature and the interlayer
tunneling strength, which we treat perturbatively. We discuss how
these results might extend to the case of true disorder.
\end{abstract}
\vskip 1cm \pacs{73.50.Jt}
\maketitle
\section{Introduction}
It has long been expected theoretically that bilayers offer the
possibility of Bose condensation of interlayer
excitons\cite{bilayer-keldysh}. While investigations in electron-hole
bilayers are ongoing\cite{bilayers-no-B}, the electron bilayer system in a
large magnetic field at total electron filling $\nu=1$ offers a
different way of realizing this goal\cite{eisenstein-macd}. There are
two important length scales in this system, the interlayer separation $d$ and the
magnetic length $l=\sqrt{\frac{\hbar}{eB}}$, (MKS units) where $e$ is
the magnitude of the electron charge and $B$ is the magnetic
field. In addition to the dimensionless tuning parameter $d/l$, there
may also be a layer imbalance $\nu_{\uparrow}-\nu_{\downarrow}$, where
$\uparrow,\ \downarrow$ refer to the top and bottom layer, and an
in-plane field $B_{\parallel}$. The discussion in this paper will be
restricted to the case of balanced layers with
$\nu_{\uparrow}=\nu_{\downarrow}=\half$ and $B_{\parallel}=0$. In this
case, by a particle-hole transformation in one layer, one can
immediately see the possibility of exciton formation and condensation
for small $d/l$.

The first theoretical works on this system two decades ago
predicted in the clean limit that at $T=0$ and small $d/l$, the system
could be thought of either as a exciton Bose condensate or as a
pseudospin quantum Hall ferromagnet\cite{cqfm} with a planar anisotropy. The
presence of a gapless, linearly dispersing, Goldstone mode (in the
absence of interlayer tunneling) was noted\cite{fertig89} indicating a
spontaneous broken symmetry, as was the expectation that when the
interlayer tunneling amplitude $h$ became nonzero but small compared
to the other energy scales, there would be a delta-function peak in
the interlayer conductance $G_{IL}$ at zero bias\cite{wen}. Upon the
application of an in-plane field $B_{\parallel}$, the peak is expected
to split into two symmetric peaks separated by a bias voltage
proportional to
$B_{\parallel}$\cite{bilayer-original-th}. Furthermore, the tunneling
is expected to occur within a Josephson length
($l_J=\sqrt{\frac{J}{h}}$, with $J$ being the pseudospin stiffness) of
the contacts\cite{wen,bilayer-original-th}. Finally, for $h=0$, a
Berezinskii-Kosterlitz-Thouless transition\cite{BKT,JKKN} at a nonzero
temperature $T_{KT}$ is expected, and the counterflow conductance,
where current flows in opposite directions in the two layers, is
expected to be infinite for $T<T_{KT}$\cite{bilayer-original-th}. An
excellent review summarizes the theoretical expectations in the clean
system\cite{multicomp-review}.

Experimentally, it is found that as $d/l$ varied, there is a
transition\cite{phase-transition-d/l} between a compressible phase
with unquantized Hall conductance for large $d/l$ (presumably
adiabatically connected to a system with two decoupled $\nu=\half$
Composite Fermion\cite{jain-cf} Fermi seas\cite{CF-FS} in the two
layers) and an incompressible phase for $d/l\le1.6$ with a quantized
Hall conductance.

Where the experimental results deviate from theoretical expectations
is in the interlayer
tunneling\cite{spielman-tunneling-drag,tiemann1,spielman-coll-mode}
and counterflow\cite{counterflow-ex} properties of the incompressible
phase.  It is never possible in an experimental sample to make $h$
precisely zero, but it can be made much smaller (of order $100\mu K$)
than any other energy scale, including the temperature. In such
samples, there is a peak at zero-bias in $G_{IL}$, but its width
remains nonzero even at the lowest
temperatures\cite{spielman-tunneling-drag}. Upon the application of an
in-plane field, features at nonzero bias are seen to vary in a linear
manner with $B_{\parallel}$, but the peak at zero-bias
remains\cite{spielman-coll-mode}, contrary to theoretical
expectations, and retains most of the weight. The counterflow
conductivity increases as $T$ decreases, but remains finite down to
the lowest measured $T$\cite{counterflow-ex}. Contrary to theoretical
expectations, interlayer tunneling occurs throughout the area of the
sample\cite{area-law} rather than within a Josephson length of the
contacts. Recently, critical interlayer tunneling currents have been
measured\cite{tiemann-area-law} (with some puzzling
aspects\cite{caution-critical}), and are also found to be proportional
to the area of the sample. Finally, while a sharp drop in the peak
value of the interlayer conductance is seen at a particular $T$, the
phenomenology of this nonzero-temperature
transition\cite{phase-tr-in-bilayer} (if indeed it is a thermodynamic
transition) does not seem to be Berezinskii-Kosterlitz-Thouless-like.

Much theoretical
work\cite{fertig89,wen,bilayer-original-th,stone96,bilayer-th1,bilayer-th2,bilayer-th3,bilayer-th4,CS-effective-th,burkov-macd,fertig2002,straley2003,lee-rapsch-chalker,abolfath-macd,sheng-wang,CFs-in-bilayers,mutual-CF,wang,huse,bilayer-phase-dia,bilayer-us,roostaei,subirme,eastham-cooper2009,su-macd}
has been carried out on this system in the past two decades in several
directions. We will not attempt to review the vast literature, except
to comment that no satisfactory fundamental explanation of the entire
spectrum of discrepancies noted above has been proposed yet. It has
been recognized that an extra mechanism of dissipation seems to be
active in this system, as seen in the finite value of the counterflow
conductivity\cite{counterflow-ex}, and in the nonzero width of the
zero-bias peak in $G_{IL}$. Assuming such a mechanism
phenomenologically, several authors have successfully described some
fraction of the experimental results, as for example in
Ref. \cite{bilayer-th2}. Recent efforts have focused on the effects of
quenched
disorder\cite{straley2003,lee-rapsch-chalker,sheng-wang,wang,huse,bilayer-us,roostaei,subirme,eastham-cooper2009},
and succeeded in explaining some aspects of the
experiments. Theoretical evidence from studies of critical
counterflow\cite{abolfath-macd} and tunneling currents\cite{su-macd}
also lead to the identification of quenched disorder as the root cause
of the discrepancies between theory and experiment.

In this paper we will start with the assumption that quenched disorder
is responsible for the qualitative differences between theory in the
clean limit and experiment. As Efros\cite{efros} showed long ago, in
any incompressible system, disorder necessarily has a nonperturbative
effect, because the clean system cannot screen linearly. In
two-dimensional electron gases, there is the further feature that the
fluctuations of the disorder potential, created by positional
flutuations of the remote dopants, diverge at long distances. In the
semiclassical Efros picture, the quantum Hall system screens by
forming compressible puddles (of typical size $s$, the distance
between the 2DEG and the dopant layer) separated by incompressible
strips of width a few magnetic lengths. This picture is supported by
imaging experiments\cite{imaging}. Based on this picture of strong
smooth disorder, two of the present authors previously
presented\cite{bilayer-us} a classical coherence network model
displaying some of the experimental phenomenology, such as the
proportionality of the interlayer tunneling conductance to the
area\cite{area-law} rather than length of the contacts. A more
detailed, though still classical, calculation\cite{eastham-cooper2009}
has recently used the coherence network model to predict a large
enhancement of the Josephson length due to disorder.

Here we take a slightly different approach, inspired by studies of the
Bose-Hubbard model\cite{bose-hubbard} and one-component quantum Hall
systems in a periodic potential\cite{plateau-lattice}. These studies
were motivated by trying to develop a field-theoretic understanding of
the superfluid-Mott Insulator transition for bosons or the plateau
transition for quantum Hall systems. Note that without a potential,
Galilean invariance means that one cannot even obtain a transition in
both these cases. Once a second-order transition has been obtained in
the presence of a periodic potential, one may examine the relevance or
irrelevance of operators pertaining to various kinds of
disorder\cite{plateau-tr-disorder} at this fixed point.

Combining ideas from the Efros picture\cite{efros} and the plateau
transition work\cite{plateau-lattice}, one may hope that a periodic
potential will capture the essential nonperturbative features of
quenched disorder, and true disorder can be added
later\cite{plateau-tr-disorder}. In this paper we will carry out the
first part of this program, ending with some speculations regarding
true quenched randomness at the end.

As a further simplification, we will assume that fermionic degrees of
freedom are gapped out and focus exclusively on the pseudospin degrees
of freedom as the only relevant ones. This also involves freezing out
the real spin, which may not be quite correct for typical experimental
samples\cite{real-spin-in-bilayer,real-spin-bilayer-th}. However,
recently it has been shown that the interlayer coherent phase is
robust to large Zeeman energies, and survives full polarization of
real spin\cite{phase-tr-ful-pol}. Our study will be of direct
relevance to that system, and the qualitative results will likely
hold for the low Zeeman energy case as well. From now on, we will
refer to ``spin'' always meaning the pseudospin vector, represented by
a unit vector $\bn$. An important ingredient is the spin-charge
relation\cite{lee-kane,shivaji-skyrmion}, which holds in
multicomponent systems in a quantum Hall
phase\cite{bilayer-original-th,multicomp-review}. In our context, this
states that slow variations of spin contain Coulomb charge:
\beq
\delta \rho(\br)=\frac{e}{4\pi} \bn\cdot\big(\partial_x\bn\times\partial_y\bn\big).
\eeq
It is through this $\delta \rho$ that the spins couple to the external
potential. For ease of computation we will put our spins on a square
lattice, with the lattice spacing chosen to be of the order of the
magnetic length $l$. In this case the above, continuum, expression has
to be replaced with the spherical area\cite{subir-af-review} subtended
by the three non-coplanar spins $\bn_1,\ \bn_2,\ \bn_3$,
\beq
\delta Q_{123}=\frac{e}{2\pi}\tan^{-1}\bigg(\frac{\bn_1\cdot\bn_2\times\bn_3}{1+\bn_1\cdot\bn_2+\bn_2\cdot\bn_3+\bn_3\cdot\bn_1}\bigg).
\label{Qtriplet}\eeq

The Hamiltonian of our model has the form
\beqr
\cH=& -J\sum\limits_{\langle\br \br'\rangle}\big(n_x(\br)n_x(\br')+n_y(\br)n_y(\br')\big)+\frac{\Gamma}{2}\sum\limits_{\br} \big(n_z(\br)\big)^2\nonumber\\
&-h\sum\limits_{\br}n_x(\br)-V_0\sum\limits_{\Box} f(X,Y)\delta Q_{\Box}+
H_U[\delta Q^2].
\label{Hschematic}\eeqr
Here $J$ is the nearest-neighbor ferromagnetic coupling, and is
related to the continuum spin stiffness, $\Gamma$ is the interlayer
charging energy, $h$ is the interlayer tunneling amplitude, and $V_0$
is the strength of the periodic potential with $f(X,Y)$ being its
specific functional form, defined on dual lattice sites ${\bf R}=(X,Y)$.
The sum in the third term, $\sum\limits_{\Box}$, refers to a sum
over plaquettes of the lattice, and
we have also introduced a local Hubbard
interaction $H_U$
to model
the energy cost (of order the interaction scale
$E_c=\frac{e^2}{\epsilon l}$) of localizing a charge within a distance
of $l$.  We will specify the precise form of these last two terms in
the next section. Most of our results are for $J=\Gamma$, with our
Hubbard $U$ being about ten times $J$, for different values of
$V_0$. We also vary the size of the unit cell between $12 \times 12$
and $16 \times 16$ lattice sites, with each representing an area of
order $100-300l^2\approx s^2$ ($s$ is the distance between the 2DEG
and the dopant layer), consistent with values of the expected size of
Efros puddles for relevant experimental samples.

Our results, which we summarize here, and elaborate upon in later
sections, show that there are some surprising features present in this
model. Consider first $h=0$. We find a sequence of ground state
transitions where the topological content of the spin configuration
changes. For sufficiently large $U$, the first transition is
second-order while the others are generically weakly
first-order. There is always a linearly dispersing Goldstone mode
(which we will henceforth call the $G$-mode) indicating the broken
$U(1)$ symmetry for $h=0$. The next higher mode is usually
quadratically dispersing (we will call it the $Q$-mode), with a gap
that sometimes becomes very small near the transitions. In the case
when the transition is continuous, the gap of the $Q$-mode vanishes at
the transition continuously. It is important to note that true
disorder will generically convert first-order transitions to
second-order\cite{RFIM}, and thus the $Q$-mode is expected to be truly
gapless at the ground state transitions in the disordered model. Two
distinct effects can be traced to the transitions and the properties
of the $Q$-mode near them. At values of $V_0$ near the transitions,
there are configurations close in energy with a different topological
density, which means that the system becomes highly polarizable. In an
effective $XY$ model (to be described in Section) this leads to a
greatly reduced core energy for vortices, or a greatly increased
fugacity, which strongly suppresses the KT-transition
temperature. Secondly, when the $Q$-mode gap is small, thermal
fluctuations of the $Q$-mode are present at realistic temperatures,
and strongly suppress the interlayer tunneling amplitude $h$. Finally,
the low-lying $Q$-excitations offer one possible microscopic mechanism
for the pervasive and puzzling dissipation seen in experiments. These
are the main results of our work.

Let us remark briefly about the realistic case $h\ne0$ but smaller than
any other energy scale in the problem, including the temperature. The
ground state phase transitions are controlled by the competition
between the core energies of merons/antimerons and the external
potential. Both these are on the scale of $E_c$. Therefore, we expect a nonzero
but tiny $h\ll T,\ V,\ E_c$ will not affect the transitions in any
qualitative way. Similarly, for the collective mode dispersions, the
primary effect of a tiny $h$ is to gap the $G$-mode at $q=0$. We
expect all other modes to be robust against the introduction of a
small $h$.

It is important to note that in the clean system, $h$ is relevant
below $T_{KT}$\cite{JKKN}, and the regimes of small and
large $h$ are adiabatically connected to each other, with no phase
transition separating them. This is not true for the system with a
periodic potential or strong, smooth disorder. As $h$ is increased
while all other parameters (such as $E_c$ and $V$) are kept fixed, the
system will undergo ground state phase transitions in which the
topological density reduces, presumably becoming the uniform
ferromagnetic state for very large $h\gg V$. This qualitative
distinction between weak and strong tunneling in the presence of a
periodic potential (and presumably a strong smooth disorder potential)
is consistent with experiments, which see the almost-superfluid
phenomenology only for weak $h$.

The plan of this paper is as follows: In Section II we will present the
precise definition of our model and present sample ground-state configurations.  Section III presents our method for computing the collective mode spectra of
excitations above the ground-state, and discusses the effects of the potential strength $V$, the inter-layer tunneling $h$, and the interaction strength $U$ on this.
This is followed by a discussion of how one may extract the spin stiffness
of our model from the numerical results.
Section IV discusses two non-trivial fluctuation effects in this system,
first how the quadratically dispersing collective mode can suppress
inter-layer tunneling, and then the suppression of $T_{KT}$ near
quantum phase transitions due to vortex excitations.
We end with some conclusions, caveats,
and speculations on what our studies tell us about the case of true
disorder in Section V.
\linebreak
\section{Model and Ground-States}
\label{groundstate}
The form of the Hamiltonian of our model, Eq. (\ref{Hschematic}), is
guided by the following considerations: (i) We ignore fermionic
excitations at the very outset, since in any incompressible state, the
energy to create a fermion is of the order of the Coulomb scale $E_c$,
which is much larger than the other relevant energy scales. Similarly
we ignore the dynamics of the real spin, assuming that it is fully
polarized\cite{phase-tr-ful-pol}. (ii) We put the system on a lattice
for computational convenience\cite{burkov-macd}. As long as all the
relevant length scales, such as the size of the puddles, are much
larger than the chosen lattice scale, we expect that our results will
be physically correct. To be specific, we will choose the lattice
spacing $a=l\sqrt{2\pi}$ such that each elementary plaquette has the
area of one cyclotron orbit, and at $\nu_T=1$, has one electron. (iii)
In the continuum, when there is no external potential, the Hamiltonian
should reduce to the continuum form\cite{bilayer-original-th} in the
long-wavelength limit. (iv) The coupling of the external potential to
the spin ({\it i.e.}, layer) degrees of freedom must be through the lattice
generalization\cite{subir-af-review} of the topological
charge\cite{lee-kane,shivaji-skyrmion}. (v) To account for
short-distance effects beyond the spin stiffness, we introduce a
Hubbard on-plaquette interaction term. (vi) Finally, we ignore the
long distance component of the
Coulomb interaction between induced charges for computational
convenience, assuming that the Hubbard term can capture all qualitative
effects.

We are now ready to make the form of the Hamiltonian precise. We
consider a square lattice of points labelled by $\br$, with the
corresponding dual lattice labelled by $\bR$, with the convention that
the site $\bR$ corresponding to $\br$ lies half a lattice unit to the
right and above $\br$. The external potential naturally lives on the
dual lattice. Each dual site $\bR$ defines a plaquette, whose sites
can be labelled $\br_1,\ \br_2,\ \br_3,\ \br_4$, starting from
$\br=\br_1$ and going counterclockwise. The topological charge
corresponding to a given triplet of spins was given above in
Eq. (\ref{Qtriplet}). The charge in a plaquette is defined as
\beq
\delta Q_{\Box}(\bR) =\delta Q_{123}+\delta Q_{134}=\delta Q_{124}+\delta Q_{234}.
\label{deltaQbox}\eeq

To fully specify the third term on the RHS of Eq. (\ref{Hschematic})
we need the functional form of $f(X,Y)$. We choose the simplest possible periodic form:
\beq
f(X,Y)=\sin\big(\frac{2\pi X}{Na}\big)\sin\big(\frac{2\pi Y}{Na}\big).
\label{f(X,Y)}\eeq
$Na$ is the size of the unit cell, which contains four puddles.  We
have experimented with other forms and observed no qualitative
differences. Finally, we turn to the last term of
Eq. (\ref{Hschematic}). In order to eliminate spurious configurations
with equal and opposite topological charges in triplets $123$ and
$134$ but no net charge in the plaquette, we write the Hubbard term as
the symmetrized sum of all the triplets in a plaquette:
\beq
H_U=\frac{U}{4}\sum\limits_{\bR} \bigg(\big(\delta Q_{123}\big)^2+\big(\delta Q_{134}\big)^2+\big(\delta Q_{124}\big)^2+\big(\delta Q_{234}\big)^2\bigg).
\label{Hubard-term}\eeq

Hartree-Fock calculations are a reasonable starting point for
estimating the values of the various parameters entering the
Hamiltonian. $J$ and $\Gamma$ are of the order of a few percent of
$E_c$ in such calculations, while $U$ should be of the order of $E_c$,
being the energy to localize a charge of order $e$ within a length
scale $l$.  We will typically assume $J=\Gamma$, and $U$ between 8 and
20 times $J$. Most of our results are for the size of the unit cell
being $N=16$, but we will show results for $N=12$ as
well. These values are realistic given $l\approx 200\AA$ and the
distance to the dopant layer $s\approx 2000\AA$.

We parameterize the spins by $z(\br)\equiv n_z(\br)$ and $\phi(\br)$
which is the $XY$ angle
[$\,\,n_x(\br)=\sqrt{1-z(\br)^2}\cos\phi(\br),\ \
n_y(\br)=\sqrt{1-z(\br)^2}\sin\phi(\br)\,\,$]. We find the ground state
configurations by starting with a random seed configuration, followed
by simulated annealing. Several different random
seeds were tried at every value of $V_0$ to eliminate the possibility
of settling into a metastable minimum.

\begin{figure}
\includegraphics[width=60mm,height=60mm]{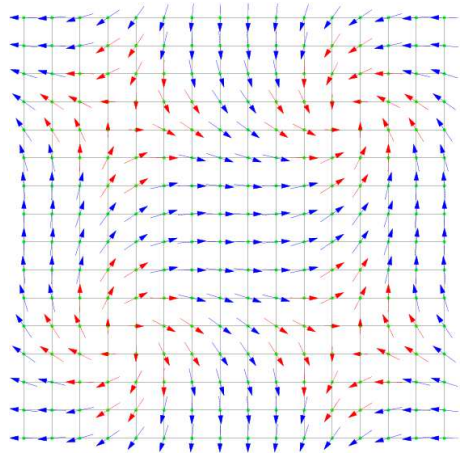}
\caption{({\it Color online}) The ground state configuration for a $16\times16$ unit cell
with the strength of the periodic potential (in units where $J=1$)
being $V=3.0$, and the Hubbard interaction is $U=8.0$. The lengths of
the spins denote their planar projection. Note a vortex/antivortex at
the center of each puddle.}
\label{figV32darrow}
\end{figure}

We begin by showing the ground state spin configuration for a $16\times16$ unit
cell at $V=3$ and $U=8$ (in units of $J=\Gamma=1$). As shown in
Fig. (\ref{figV32darrow}), there is a vortex/antivortex at the center
of each puddle, partially screening the external potential. The length
of the arrow indicates the projection of the spin in the $xy$-plane,
while the color indicates whether the spin points in the up or down
direction in $z$.
%
%
A different way to plot the same configuration is to look instead at
the topological density of the spins. This is shown as a three-dimensional plot in
Fig. (\ref{figV33dtopo}).
%
%
%
\begin{psfrags}
\begin{figure}
\includegraphics[scale=1.0]{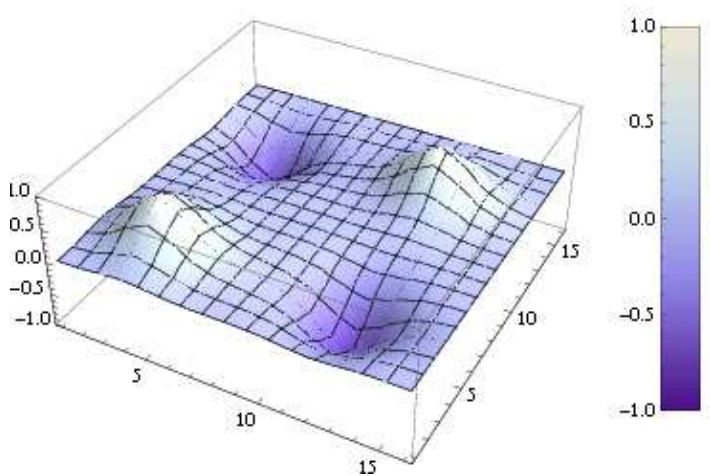}
\caption{({\it Color online})The topological density of the spins in the ground state configuration for a $16\times16$ unit cell
with $V=3.0$, and $U=8.0$ shown as a three-dimensional plot.}
\label{figV33dtopo}
\end{figure}
\end{psfrags}
A more complicated configuration is shown in Fig. (\ref{figV72darrow}), where $V$ has now been
increased to $7$. Each puddle now has two vortex-antivortex pairs.
\begin{psfrags}
\begin{figure}
\includegraphics[scale=0.4]{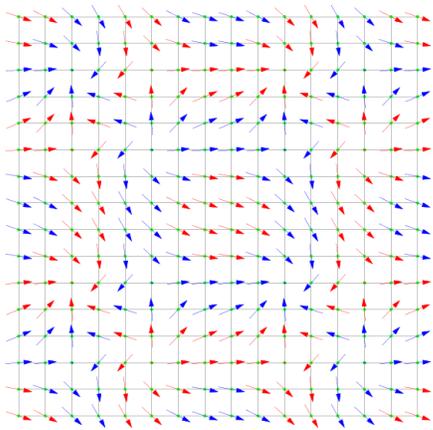}
\caption{({\it Color online}) The ground state configuration for $V=7.0$ and $U=8.0$.}
\label{figV72darrow}
\end{figure}
\end{psfrags}
The corresponding topological density is plotted as a countour plot in
Fig. (\ref{figV72dtopo}).
\begin{psfrags}
\begin{figure}
\includegraphics[scale=1.0]{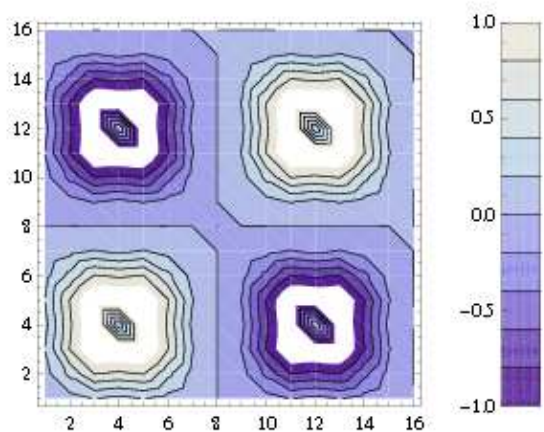}
\caption{({\it Color online}) The topological charge in the ground state configuration for $V=7.0$ and $U=8.0$.}
\label{figV72dtopo}
\end{figure}
\end{psfrags}
The phase transitions correspond to values of $V$ where the ground
state energy of one configuration becomes lower than another. The very
first of these transitions (for sufficiently large $U$) is
second-order. For small $V$ the ground state is uniform and
ferromagnetic as at $V=0$. At a critical $V=V_c$ (which happens to be
$V_c=4.2$ for $U=18$) there is an instability of the ground state,
which, when followed, leads to a ground state with a nonzero
topological density. While the very first of these transitions is
second-order, the rest seem to generically be weakly first-order.  The
ground state energies cross with different slopes as functions of $V$.

In Fig. (\ref{fig12by12}) we show the ground state of the $12\times12$
unit cell at $V=3$ and $U=8$. The presence of two vortex-antivortex
pairs is evident, as is the qualitative similarity to the case of the
$16\times16$ unit cell of Fig. (\ref{figV32darrow}). Generally we find
that all the qualitative features of the different ground states are
the same for different sizes of the unit cell, but the specific values
at which the transitions happen do depend on unit cell size.
\begin{psfrags}
\begin{figure}
\includegraphics[scale=0.40]{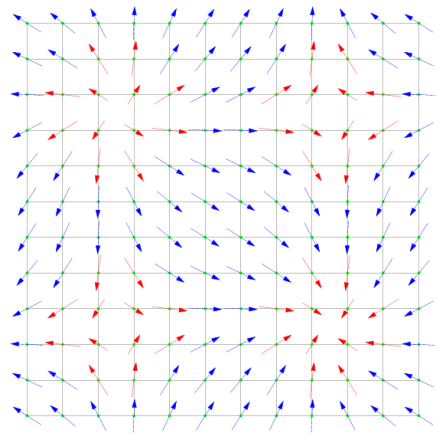}
\caption{({\it Color online}) The configuration of the $12\times12$ lattice at $V=3$ and $U=8$.
Comparing to Fig. (\ref{figV32darrow}) we see that the presence of the
vortex-antivortex pairs is independent of the size of the unit cell,
though the values of critical $V$ at which the ground state
transitions take place are different. }
\label{fig12by12}
\end{figure}
\end{psfrags}

\section{Deviations from the Ground State}

Having obtained classical groundstates of this system, we are in
a position to consider the energetics and dynamics associated
with deviations from it.  In what follows we will focus on
small deviations, such that the system can be analyzed within
a quadratic theory.  Two types of quantities are of particular
interest, the collective modes of the system, and the effective
spin stiffness.  The first of these can be used to explore the
effects of thermal fluctuations, and the second is important
in determining the effects of thermally generated vortices
({\it i.e.}, Kosterlitz-Thouless physics.)

\subsection{Collective Modes}

Once a periodic ground state configuration $\zb(\br),\ \phib(\br)$
has been found, one can look at collective mode dynamics around this
state. For dynamics, one needs the full action, which is
\beq
S=S_B-\int dt \cH,
\eeq
where $S_B$, the Berry phase term\cite{haldane-berry}, measures the
spherical area covered by a closed path followed by a spin in time,
\beq
S_B=\frac{\hbar}{2}\int dt\int\limits_{0}^{1}d\xi \sum_{\br}\bn(\br,t,\xi)\cdot\partial_t\bn(\br,t,\xi)\times\partial_{\xi}\bn(\br,t,\xi).
\label{Berryphaseterm}\eeq
Here $\xi$ is an auxiliary variable and $\bn(\br,t,\xi)$ is chosen such
that at $\bn(\br,t,0)={\hat{e}_z}$ and
$\bn(\br,t,1)\equiv\bn(t,\br)$ \cite{stone96}. Also, we have made the $\hbar$
explicit, and assumed that the spin is $\half$. This is reasonable
since the underlying fermionic system has one electron per
cyclotron orbit; with our choice of lattice spacing this translates to
one electron per site.

The Berry phase term may be conveniently expressed as
\beq
S_B=\int dt\half \sum\limits_{\br}\big(z(\br)\dotphi(\br)-\phi(\br)\dotz(\br)\big).
\label{SB-zphi}\eeq
Following standard methods, and recalling that any variation of $\bn$
has to be perpendicular to $\bn$ we obtain the equation of motion
\beq
\partial_t\bn=\bn\times\frac{\delta \cH}{\delta \bn}.
\eeq
The ground state configuration is static, and satisfies
\beq
\frac{\partial \cH}{\partial z(\br)}|_{\zb,\phib}=\frac{\partial \cH}{\partial \phi(\br)}|_{\zb,\phib}=0.
\eeq
Expanding in small fluctuations around the ground state,
\beq
z(\br)=\zb(\br)+\zeta(\br),\ \ \ \phi(\br)=\phib(\br)+\vphi(\br),
\eeq
we obtain the linearized equations of motion
\beqr
\dotzeta(\br)=&-\frac{\partial^2\cH}{\partial \phi(\br)\partial z(\br')}\zeta(\br')-\frac{\partial^2\cH}{\partial \phi(\br)\partial \phi(\br')}\vphi(\br')\nonumber\\
\dotvphi(\br)=&\frac{\partial^2\cH}{\partial z(\br)\partial z(\br')}\zeta(\br')+\frac{\partial^2\cH}{\partial z(\br)\partial \phi(\br')}\vphi(\br').
\eeqr
As can be seen, the Hessian matrix of second derivatives (computed at
the ground state configuration) appears on the right hand side,
\beq
H_{ss'}(\br,\br')=\frac{\partial^2 \cH}{\partial w(\br,s)\partial w(\br',s')},
\eeq
where we have introduced indices $s,\ s'$ taking the values $1,\ 2$
for $\zeta$ and $\vphi$ respectively, thereby organizing
$\zeta(\br)=w(\br,1),\ \vphi(\br)=w(\br,2)$ into a two-component
vector. It is important to note that for $h=0$ one can rotate all the
$\phi$ by the same amount producing no change in energy, which means
that the Hessian matrix $H$ has an eigenvector with a zero eigenvalue.

The next step is to exploit the larger unit cell and partially
diagonalize the system in terms of a wave-vector $\bk$ lying in the
Brillouin Zone (BZ), arranging the $2N^2$ variables in a unit cell
into a single vector. We can now define the $2N^2\times2N^2$ matrices
\beqr
\tH_{ss'}(\bk:\br,\br')=&e^{-i\bk\cdot(\br-\br')}H_{ss'}(\br,\br'),\\
K_{ss'}(\br,\br')=&(\sigma_2)_{ss'}\otimes \delta_{\br,\br'},
\eeqr
where $\sigma_2$ is a Pauli matrix.
It is henceforth understood that in all matrices $\br,\ \ \br'$ are
restricted to a single unit cell. Note that both matrices are
Hermitian, but differ under transposition:
\beqr
\big(\tH(\bk)\big)^{T}=&\tH(-\bk)\nonumber\\
K^T=&-K.
\label{realityofth}\eeqr
The earlier condition on the Hessian having an eigenvector
with zero eigenvalue translates to the vanishing of the smallest
eigenvalue of $\tH$ as $\bk\to0$. This will be important for the
computation of the spin stiffness.

Now the equations of motion can be written in a compact form
\beq
\omega K \psi^{[R]}(\bk)=\tH(\bk) \psi^{[R]}(\bk).
\eeq
This is a generalized eigenvalue problem with both $K$ and $\tH$
being Hermitian. We first solve for the eigenvalues and eigenvectors
of the auxiliary problem
\beq
K\tH \psi^{[R]}_{\lam}(\bk)=\omega_{\lam}(\bk)\psi^{[R]}_{\lam}(\bk)
\label{eigvaleqn}\eeq
to get the right eigenvectors $\psi^{[R]}_{\lam}$.  Taking the Hermitian adjoint and
carrying out a few manipulations, one obtains the left eigenvectors
\beq
\bigg(\psi^{[R]}_{\lam}(\bk)\bigg)^{\dagger} K=\psi^{[L]}_{\lam}(\bk).
\label{psiL}\eeq
Using the hermiticity of $K$ and $\tH$, and
Eqs. (\ref{realityofth}) and (\ref{psiL}) it is easy to show that the
eigenvalues occur in pairs with equal magnitude and opposite sign.

The
normalization condition on the eigenvectors can be chosen to be
\beq
\bigg(\psi^{[R]}_{\lam}(\bk)\bigg)^{\dagger}K\psi^{[R]}_{\lam'}(\bk)=sgn(\omega_{\lam}(\bk))\delta_{\lam\lam'},
\label{norm}\eeq
where the $sgn(\omega_{\lam})$ is essential because $K$ is not
positive definite. Now, as a consequence of
Eqs. (\ref{eigvaleqn}) and (\ref{norm}) the eigenvectors also satisfy
\beq
\bigg(\psi^{[R]}_{\lam}(\bk)\bigg)^{\dagger}\tH\psi^{[R]}_{\lam'}(\bk)=
|\omega_{\lam}(\bk)|\delta_{\lam\lam'}.
\label{hnorm}\eeq
Henceforth we will use only the right eigenvectors, dropping the
superscript $[R]$.

We find the following results for the collective mode dispersions as functions
of $V$. Since we set the interlayer tunneling $h=0$, there is always a
gapless, linearly dispersing, Goldstone mode arising from the breaking
of the continuous $U(1)$ global symmetry. We will henceforth call this
the $G$-mode. The $G$-mode is generically the lowest mode throughout
the BZ, but may not be so close to a transition. The next higher
energy mode is quadratically dispersing at $\bq=0$, and will play an
important role in what follows. We will henceforth call it the
$Q$-mode. At the first transition, which is second-order for large
$U$, the gap of this mode vanishes at $\bq=0$. In
Fig. (\ref{V4p3U18SW}) we show the dispersions of the two lowest
energy modes at $V=4.2$ and $U=24$, close to the transition. It is
clear that the $Q$-mode is the lowest energy mode for a substantial
part of the BZ.
\begin{psfrags}
\begin{figure}
\includegraphics[scale=0.35]{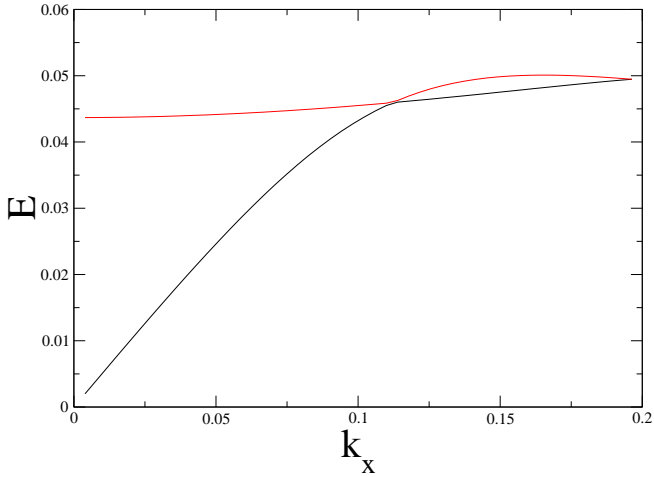}
\caption{({\it Color online}) The two lowest collective modes for $V=4.3$ and $U=18$.
This is just before the transition from the uniform ferromagnetic
state to a striped state, and is generically second-order at
sufficiently large $U$. Note the almost gapless quadratically
dispersing mode.}
\label{V4p3U18SW}
\end{figure}
\end{psfrags}

At a generic transition, however, the gap of the $Q$-mode at $\bq=0$
does not vanish. Summarizing the behavior of the $Q$-mode over the
entire range of $V$ we investigated, we present
Fig. (\ref{OmegaQvsV}), which shows the $Q$-mode gap as a function of
$V$ for $U=8$.
\begin{psfrags}
\begin{figure}
\includegraphics[scale=0.35]{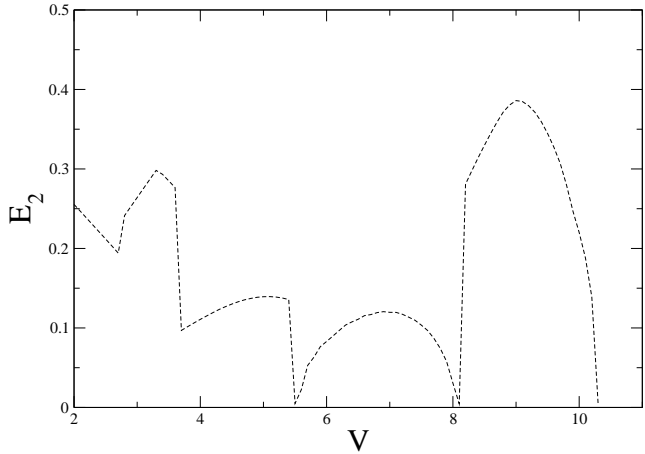}
\caption{The gap at $\bq=0$ of the $Q$-mode as a function of $V$.
Note that it does not generically vanish at a transition, though there
are transitions where it becomes particularly small. }
\label{OmegaQvsV}
\end{figure}
\end{psfrags}
We also examined the ``wave functions'' of the $Q$-mode near the
transitions.  The $\phi$-component appears to be strongly peaked at
the vortex/antivortex cores, while the $z$-component is more broadly
distributed.  Roughly speaking this may be understood in terms of
motion of the merons around their minimum potential sites, with a very
flat effective curvature, indicating that ``room'' for further merons
is developing in the potential well.  (Similar behavior has been
observed in simulations of this system using an $XY$
model \cite{straley2003}.)  Moreover, the $Q$-mode dispersion is
typically quite flat, indicating that these are fairly well-localized
excitations.
\begin{psfrags}
\begin{figure}
\includegraphics[scale=0.35]{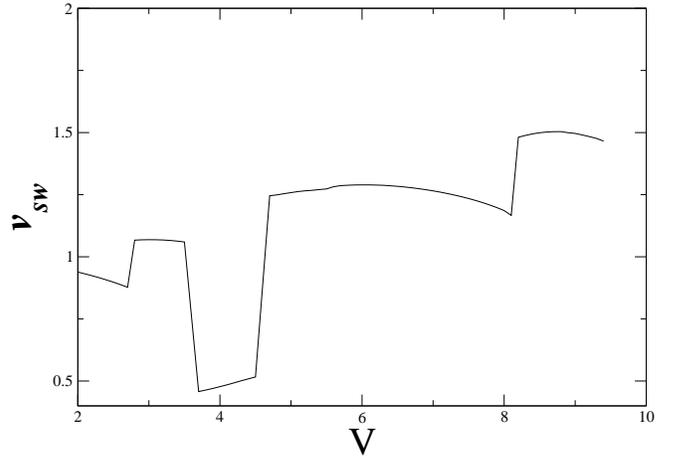}
\caption{The spinwave velocity of the  $G$-mode as a function of $V$.
Note that it never vanishes, but that it can vary by a factor of
three, and have significant discontinuities at transitions.}
\label{sw-velocity-vs-V}
\end{figure}
\end{psfrags}
Another interesting point to note, shown in
Fig. (\ref{sw-velocity-vs-V}), is that there are large fluctuations of
the $G$-mode velocity as $V$ changes. It should be kept in mind that
these are results to quadratic order in the fluctuations around the
ground state configuration. Interactions between the $G$- and $Q$-
modes could affect the gap of the $Q$-mode and the spin-wave velocity
as well.

\subsection{Ground-State Spin Stiffness}
\label{stiffness-section}
Another related quantity of interest for this system is the spin stiffness,
which determines among other things the interactions between highly
separated vortex-antivortex pairs.  We begin by describing
how it may be computed for an arbitrary ground
state configuration. We define the spin stiffness by calculating the energy $U$
for a particular spatially dependent configuration of $\vphi$ and
comparing it to the reference energy $U_{cont}$ of the same
configuration in a free continuum model,
\beq
U_{cont}=\half \cK_s\int d^2\br \big(\nabla\phi\big)^2.
\label{rhos-defn}\eeq
The least expensive way to twist the system is controlled by the
smallest eigenvalue of the Hessian $\tH(\bk)$ as $|\bk|\to0$. Note
that these eigenvalues are not the same as the energies of the
eigenmodes found in Eq. (\ref{eigvaleqn}), which are the solutions to
the dynamical problem. We consider a sample with lengths $L_x,\ L_y$
in the $x$ and $y$ directions respectively. We denote the eigenvalues of
$\tH(\bk)$ as $\e_n(\bk)$ and the corresponding eigenvectors as
$\chi_n(\bk,\br,s)$. (Here $s=1,2$ correspond respectively to fluctuations of
$\zeta$ and $\varphi$.) Since $\tH$ is Hermitian and represents a stable
ground state, the eigenvalues are all positive semidefinite, and the
eigenvectors are normalized in the usual way,
\beq
\big(\chi_n(\bk)\big)^{\dagger}\chi_{n'}(\bk)=\delta_{nn'}.
\eeq
We can express a generic fluctuation $\vphi$ of the phase field
$\phi({\bf r})$ from the ground state at
any $\br$ inside the unit cell in the presence of a unit amplitude of
the lowest eigenvector, which we call $\chi_0(\bk)$, as
\beq
\vphi(\br)=\frac{1}{\sqrt{N_c}}\sum\limits_{\bk} \sum\limits_{n} e^{i\bk\cdot\br} p_n(\bk) \chi_n(\bk,\br,2),
\label{vphigeneric}\eeq
where $N_c=\frac{L_xL_y}{N^2a^2}$ is the number of unit cells.  The reality
of $\zeta,\ \ \vphi$ forces $p_n^*(\bk)=p_n(-\bk)$ with the convention
that $\chi_n^*(\bk)=\chi_n(-\bk)$. The energy of such a fluctuation is easily seen to be
\beq
U=\sum\limits_{\bk,n}\e_n(\bk) |p_n(\bk)|^2.
\label{energy-generic}\eeq
To make the calculation convenient, we choose $\bk_{min}=k_{x,min}\he_x$ with
$k_{x,min}=\frac{2\pi}{L_x}$ being the minimum nonzero value allowed. Then, bearing in mind the reality of $\vphi$,
we can express this as
\beq
\vphi(\br)=\frac{2}{\sqrt{N_c}} p_0(\bk_{min})\chi_0(\bk_{min},\br,2)\sin\big(\frac{2\pi x}{L_x}\big).
\label{testvphi}
\eeq
Comparing the energy of Eq. (\ref{energy-generic}) to that of the
continuum version Eq. (\ref{rhos-defn}), we obtain
\beq
\cK_s=\frac{\e_0(\bk_{min})}{N^2a^2k_{min}^2|\chi_0(\bk_{min},\br=0,2)|^2}.
\label{rhos-final}\eeq
To be precise, this is the spin stiffness computed at $T=0$, because
thermal fluctuations have not been included.

The results for the spin stiffness (Fig.~\ref{stiffness}) as a function of potential strength $V$ show prononounced jumps.  The discontinuities in $\cK_s$ coincide with transitions in the topological density of the ground-state configuration.  Notice that the spin stiffness is generally of the order of $J$, the nearest-neigbor ferromagnetic coupling in Eq. \ref{Hschematic}.  Depending on the topological density in the ground-state configuration, $\cK_s/J$ can be either significantly enhanced (for example $V/J\le 8.15$) or reduced (for example, $3.75\le V/J \le 5.5$).
\begin{psfrags}
\begin{figure}
\includegraphics[scale=0.35]{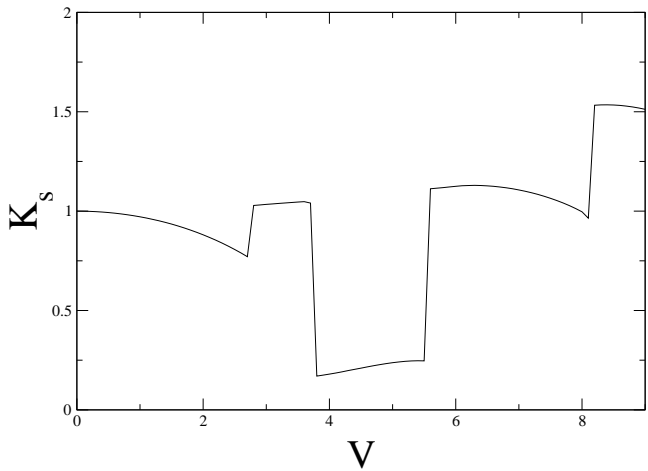}
\caption{Spin stiffness as a function of potential strength $V$, normalized to nearest-neighbor ferromagnetic exchange $J$.}
\label{stiffness}
\end{figure}
\end{psfrags}
\section{Fluctuation Effects}

In a purely quadratic theory, fluctuations will have little effect on most
quantities of interest.  However, the Hamiltonian of our system contains non-linear couplings
which can have important qualitative effects.  Already included in
our groundstate analysis is the underlying ${\cal O}(3)$ nature of the
spins, which supports the (charged) merons that are induced
by the periodic potential.  Thermal fluctuations in which
meron-antimeron pairs are generated above the groundstate can
spoil the spin stiffness of the system, rendering it dissipative,
above an effective Kosterlitz-Thouless (KT) transition temperature.
In principle, interlayer tunneling may have even more
profound effects, for example removing the possibility of
a true thermodynamic KT transition in the clean limit \cite{JKKN,fertig2002}.
In this section we
describe how fluctuation effects in this model can greatly
suppress the impact of interlayer tunneling, and furthermore
lead to a lowering of the
KT transition temperature, particularly near transitions
between different groundstates.

\subsection{Suppression of Interlayer Tunneling by Collective Modes}

We begin with an evaluation of the effective ({\it i.e.}, renormalized)
interlayer tunneling amplitude due to
quadratic fluctuations around the ground state.  As discussed
in the Introduction, this can be greatly depressed by the presence
of the $Q$-mode, particularly if the gap is quite small, as
sometimes happens in the vicinity of a groundstate transition.

 The fluctuations around the ground state can be
expanded as
\beq
w(\br,s,t)={1\over\sqrt{N_c}}\sum_{\bk,\lam} \tw_{\lam}(\bk)e^{i(\bk\cdot\br-\omega_{\lam}t)}\psi_{\lam}(\bk,\br,s),
\eeq
where $N_c$ is the number of unit cells in the lattice and
$\tw_{\lam}(\bk)$ is the amplitude of $w$ in the mode
$\lam,\bk$. The reality of $w$ means that
$(\psi_{\lam}(\bk))^*$ is an eigenvector of $\tH(-\bk)$ with
eigenvalue $-\omega_{\lam}(\bk)$, that is
\beqr
\bigg(\psi_{\lam}(\bk,\br,s)\bigg)^*=&\psi_{-\lam}(-\bk,\br,s),\\
\big(\tw_{\lam}(\bk)\big)^*=&\tw_{-\lam}(-\bk).
\label{reality1}\eeqr
Now we are ready to re-express the action in terms of these modes. In
order to carry out thermal averages, we use the imaginary time
path integral
\beqr
-S=&\int\limits_{0}^{\beta} dt \bigg(\sum_{\br} {i\over2}[\zeta(\br)\dotvphi(\br)-\vphi(\br)\dotzeta(\br)]\\
&-\half\sum_{\br\br',ss'} w(\br,s)H_{ss'}(\br,\br')w(\br',s')\bigg).
\eeqr
When we expand in normal modes, $\tw$ now acquires a dependence on $i\omega_n$ as well:
\beqr
w(\br,t)=&{1\over\beta\sqrt{N_c}}\sum\limits_{i\omega_n,\bk,\lam} \tw_{\lam}(\bk,i\omega_n)e^{i(\bk\cdot\br-\omega_n t)} \psi_{\lam}(\bk,\br),\\
&\big(\tw_{\lam}(\bk,i\omega_n)\big)^*=\tw_{-\lam}(-\bk,-i\omega_n).
\label{reality2}\eeqr
The last condition means that we can take
$\tw_{\lam>0}(\bk,i\omega_n)$ over the entire BZ and for
all $i\omega_n$ to be independent, while the $\tw_{\lam<0}$ are
dependent. Substituting the expansion for $\delta w$ in terms of the
normal modes, and using the properties of the wavefunctions
Eqs. (\ref{norm},\ref{hnorm},\ref{reality1}) we obtain
\beq
-S={1\over\beta}\sum\limits_{i\omega_n,\bk,\lam>0}|\tw(i\omega_n,\bk,\lam)|^2 \bigg[i\omega_n-\omega_{\lam}(\bk)\bigg].
\eeq
As can be seen, the real part of $-S$ is negative definite, as it
should be for the convergence of the path integral.
This simple form for the action
leads to
\beq
\langle \tw_{\lam}(\bk,i\omega_n)\tw_{\lam'}(\bk',i\omega_n')\rangle={\delta_{\lam,-\lam'}\delta_{\bk,-\bk'}\delta_{\omega_n,-\omega_n'}\over i\omega_n-|\omega_{\lam}(\bk)|}.
\eeq
Now we are ready to find the correlation functions of the fluctuations
$w$. The simplest is $<(w(\br))^2>$. Clearly, this will be
\beq \langle(w(\br,s))^2\rangle=-{1\over
N_c\beta}\sum\limits_{i\omega_n,\bk,\lam>0}{|\psi_{\lam}(\bk,\br,s)|^2\over
i\omega_n-\omega_{\lam}(\bk)}. \eeq
The sum over $i\omega_n$ can be done by standard methods to get
$n_{\lam}(\bk)+\half$, where
$n_{\lam}(\bk)=(\exp{\beta\omega_{\lam}(\bk)}-1)^{-1}$ is the boson
occupation function, so that
\beq
\langle(w(\br,s))^2\rangle={1\over N_c}\sum\limits_{\bk,\lam>0}|\psi_{\lam}(\bk,\br,s)|^2 [n_{\lam}(\bk)+\half].
\label{avesqr}\eeq
%

We now consider the effect of these fluctuations on interlayer tunneling.
The effect as we shall see is most pronounced where
the $Q$-mode gap is
small.  In our periodic model this occurs at specific values of $V$;
in a real disordered sample, there will always be regions where
this is true. We add to our original Hamiltonian a small interlayer tunnelling term,
which may be written in the form
\beq
H_{ILT}=-h\sum\limits_{\br}\sqrt{1-z(\br)^2} \cos{\phi(\br)}.
\eeq
We approximate the $Q$-mode dispersion as
\beq
\omega_{Q}(\bk)=E_{Q0}+\alpha \bk^2.
\label{Qmode-disp}\eeq
We now consider fluctuations of $\phi(\br)$ which consist of two low energy parts,
one controlled by the
$G$-mode and the other controlled by the $Q$-mode, under the assumption
that $T$ is much smaller than the energies of the other modes.  We thus write
\beq
\phi(\br)=\phi_G(\br)+\phi_Q(\br).
\eeq
We then integrate out the $Q$-mode to obtain
a ``renormalized'' interlayer tunneling term,
\beq
H_{ILT,R}=-h\sum\limits_{\br}\sqrt{1-z(\br)^2} \cos{\phi_G(\br)}e^{-\langle\phi_Q(\br)^2\rangle/2}.
\eeq
Using Eqs. (\ref{avesqr}) and (\ref{Qmode-disp}), we rewrite the last exponential
as
\beq
\exp{\int\limits_{BZ}\frac{d^2k}{(2\pi)^2} |\psi_{Q}(\bk,\br,2)|^2(n_{Q}(\bk)+\half)}.
\eeq
For $T\gg E_{Q0}$ one can approximate the occupation number of the
$Q$-mode as
\beq
n_{Q}(\bk)\approx \frac{T}{E_{Q0}+\alpha \bk^2},
\eeq
so that as $E_{Q0}\to0$, one obtains a logarithmic divergence in the $\bk$
integral, provided $\psi_{Q}(\bk,\br,2)$ remains nonzero as
$\bk\to0$. We have verified that indeed it does, leading to a
renormalization of $h$ of the form
\beq
h_R\approx h e^{-Tl^2\log{\frac{\Lambda}{{E_{Q0}}}}/\alpha}.
\eeq
Thus, in situations where ${E_{Q0}}\to0$, $h$ is strongly suppressed.

\subsection{Suppression of $T_{KT}$}
%

%
%

%
Recall that at a generic transition in our periodic potential model
${E_{Q0}}$ does not vanish. However, as we will now show, the
Berezinskii-Kosterlitz-Thouless transition temperature $T_{KT}$ separating the low-temperature, quantum Hall ferromagnet phase from the high-temperature, paramagnetic phase is
nevertheless strongly suppressed near transitions. Physically, this is
because there are two local minima with nearly the same energy, but
different topological density. This makes it very easy for the system
to screen vortex/antivortex charge, leading to a very small core
$E_{c}$ energy for vortices.

To see this quantitatively, we introduce and analyze an effective, two-dimensional Coulomb gas description of the bilayer quantum Hall ferromagnet.  At finite temperature $T$, we approximate the partition function of the quantum Hall ferromagnet $Z\approx\sum_{\{m(\bR\}} \exp(-E[\{m(\bR\}]/k_BT)$ as a sum over the positions of vortices $\{m(\bR\}$ with a Coulomb gas energy functional
\beqr
E[\{m(\bR\}]=&\half\sum\limits_{\bR\ne\bR'}\cK_s m(\bR)m(\bR')\log\big(\frac{|\bR-\bR'|}{\xi}\big)
\nonumber\\
&-E_c(V)\sum\limits_{\bR}(m(\bR))^2,
\label{kt}
\eeqr
where $\cK_s$ is the bare spin-stiffness computed in Section \ref{stiffness-section}, $m(\bR)$ is the vorticity of a
plaquette at position $\bR$, $\xi$ is an ultraviolet cutoff
scale, and $E_c$ is the core energy of each vortex/antivortex, whose
value depends on $V$, the potential strength, through the potential-dependent details of meron core structure.  

To estimate $E_c(V)$, we compare the
$T=0$ ground-state energy (computed in Section \ref{groundstate}) of the two competing states whose energies cross at the critical potential strength $V_c$.  In the simplest case State 1 is a topologically trivial configuration with no vortices, and State 2 has two $+$ and two $-$ vortices in a checkerboard pattern in the unit cell. (See, for example, the configuration shown in Fig. \ref{figV32darrow}).  Generically, they cross with some slope, with the
energy difference per site being
\beq
\Delta E=E_2-E_1=\alpha (V-V_c),
\label{alpha}
\eeq
where $V_c$ is the transition point. We obtain $\alpha\approx 0.12$ numerically by
comparing the energies of the ground and metastable states near
$V_c$.
Without loss of generality, we choose the vortex core size $\xi$ in (\ref{kt}) such
that the vortex core energy vanishes as
one approaches the transition.   Together with (\ref{alpha}), this choice implies the following form for the vortex core energy in the vicinity of the transition at $V_c$:
\beq
E_c\approx \frac{\alpha}{4}|V-V_c|.
\label{core-energy-estimate}\eeq
Note that as one passes through the transition, the high and low energy
states interchange roles, and (\ref{kt}) still governs the fluctuations
of the system for $V<V_c$, with the ``absence of a vortex'' playing
the role of vortices in the free-energy functional.  Thus 
the vortex core energy $E_c(V)$ is non-negative across the transition at $V_c$.

%
\begin{psfrags}
\begin{figure}
\includegraphics[scale=0.35]{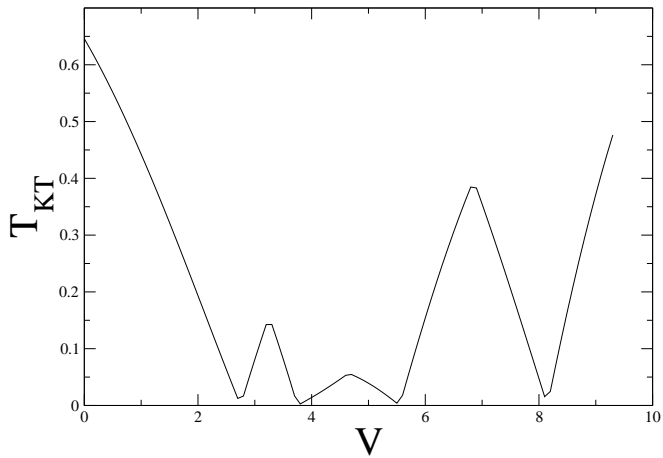}
\caption{Finite temperature phase diagram of the bilayer quantum Hall ferromagnet in a periodic potential $V$ (Measured in units of $J$ the nearest-neighbor ferromagnetic exchange energy, using a $16\ell_B\times16\ell_B$ unit cell with local Coulomb repulsion energy $U=8J$).  Dramatic decreases in the transition temperature $T_{KT}$ (measured in units of $J/k_B$) separating the low-temperature ferromagnetic phase from high-temperature paramagnetic phase occur due to changes in the topological density of the ground-state at critical potential strengths.}
\label{kt-temp}
\end{figure}
\end{psfrags}
Following the standard analysis \cite{JKKN} of the two-dimensional Coulomb gas (\ref{kt}), 
we find an implicit equation for $T_{KT}$ in terms of the vortex core energy $E_c$ and 
the ground-state spin stiffness $\cK_s$,
\begin{equation}
E_c\approx k_B T_{KT} \left | \ln \left [ \frac{1}{\sqrt{2} \pi} \left(1- \frac{2k_B T_{KT}}{\pi \cK_s}\right )\right ] \right |.
\label{separatrix}
\end{equation}

We plot in Fig. (\ref{kt-temp}) the result of numerically solving (\ref{core-energy-estimate}) and (\ref{separatrix}) for $T_{KT}$ as a function of the potential strength $V$.  This is one of the central results of this
paper.  Note that typically, $T_{KT}$ is suppressed by an order of
magnitude compared to the Hartree-Fock predictions for a bilayer quantum Hall ferromagnet with no periodic potential.

Near the ground-state transitions, the sharp reduction in $E_c$, the core energy for vortex excitations, described by  (\ref{core-energy-estimate}), is the principal cause of the dramatic suppression of the low-temperature quantum Hall ferromagnet phase.  Far from the ground-state transitions, a secondary effect of the periodic potential on the phase diagram arises through the ground-state spin stiffness $\cK_s$: $k_BT_{KT}\rightarrow \pi\cK_s/2$ for $|V-V_c|\gg 2\pi\cK_s/\alpha$.  As shown in Fig. \ref{stiffness}, the periodic potential reduces $\cK_s$ dramatically for some values of the potential strength $V$, leading to a suppression of $T_{KT}$ compared to the transition temperature in the absence of a periodic potential.

The Coulomb gas description (\ref{kt}) is expected to over-estimate $T_{KT}$ since it contains only a single type of vortex, and focuses on the spin configurations (States 1 and 2) with the lowest $T=0$ energy.  In the critical regime $V\approx V_c$ we observe numerically that the $T=0$ energy of other states approach that of States 1 and 2.  Describing the effects of these other states would require generalizing (\ref{kt}) to consider multiple types of vortices.  On general grounds, one expects that including more types of vortex fluctuations reduces the finite-temperature stability of the bilayer quantum Hall ferromagnet phase, so the result reported in Fig. \ref{kt-temp} is expected to be an {\it upper} bound on $T_{KT}$.

\section{Conclusions, Caveats, and Open Questions}
It has been clear for some time that there are qualitative
discrepancies between theoretical predictions for clean systems and
the actual phenomenology of $\nu=1$ quantum Hall bilayers at small
layer separation. There is a wide consensus in the community that
these discrepancies are due to quenched disorder.

Our goal in this paper is to investigate some of the nonperturbative
effects of quenched disorder in balanced $\nu=1$ quantum Hall
bilayers. Our principal premises are: (i) A strong periodic potential
can mimic some of the nonperturbative effects of disorder, and (ii)
one needs to focus only on the (pseudo)spin physics. While these
premises (especially (i)) can be debated in the context of bilayer
quantum Hall systems, there is ample historical evidence of the
fruitfulness of (i) in the Bose-Hubbard model and the problem of the
quantum Hall plateau transition. Other limitations of this work
include the neglect of long-range Coulomb interactions between induced
topological densities, the neglect of the renormalization of the
classical ground states we found due to fluctuations, the neglect of
interactions between the low-lying modes, especially the $G$- and
$Q$-modes, and the perturbative treatment of the interlayer
tunneling. We see this work as a first step in a systematic
investigation of these effects.

Starting from the (rather minimal) model Hamiltonian,
Eq. (\ref{Hschematic}), we study the ground states and the collective
excitations numerically for interlayer tunneling $h=0$. As the
strength of the periodic potential increases, we observe generically
first-order transitions between states with different topological
densities. Occasionally, these transitions are weakly first-order,
with a new, charge-carrying, quadratically dispersing mode becoming
nearly gapless at the transition. Such a mode can suppress the
interlayer tunneling amplitude strongly at nonzero $T$.  Even when the
transitions are strongly first-order, we show that vortices become
very easy to create, and drive the Berezinskii-Kosterlitz-Thouless
transition temperature to zero at the transition.

It is important to note that there is a qualitative difference between
weak and strong $h$ in our approach. While our numerics are carried
out only for $h=0$, the inclusion of an interlayer tunneling amplitude
much smaller than any other energy scale will not affect any of the
qualitative physics we uncover. However, if the strength of the
interlayer tunneling $h$ is increased while other parameters such as
$E_c$ and the strength of the periodic potential $V$ are kept fixed,
we expect the system to undergo a set of ground state phase
transitions leading ultimately to the uniform ferromagnetic ground
state at very large $h$. This qualitative distinction between weak and
strong tunneling does not exist in the clean model\cite{JKKN}, but is
consistent with experiments.

Based on our results for the periodic potential, we can speculate
about the effects of true quenched disorder. There are several
important effects: Firstly, it can be shown via a mapping to the
random field Ising model\cite{RFIM} that disorder converts the
sequence of first-order transitions we found as $V$ increases into a
sequence of second-order transitions. While this mapping is rigorously
provable only for classical models, it is believed to hold for many
quantum phase transitions as well\cite{random-quant,subir-book} (the
exceptions seem to be transitions that are completely smeared out and
destroyed by disorder\cite{smeared-by-disorder}). This sequence of
second-order transitions maintains the qualitative distinction between
weak and strong tunneling mentioned earlier. Furthermore, at the
quantum level, this implies that the $Q$-mode {\it generically}
becomes gapless at ground state transitions with true
disorder. Secondly, even away from a ground-state transition, in a
system with quenched disorder there will be large regions which are
close to a transition (the system is in a Griffiths
phase\cite{griffiths}).
Being in a Griffiths phase also means that
excitations of arbitrarily low energy are available fromlarge rare
regions close to the transition, which leads to divergent
low-frequency susceptibilities throughout the Griffiths
phase\cite{div-suscep-in-griffiths} at $T=0$.  For $T\ne 0$, since
vortices are easy to create in such regions, $T_{KT}$ is expected to
be strongly suppressed throughout the sample. This is consistent with
the fact that no low-temperature phase with the phenomenology of the
BKT power-law phase (including a Josephson-like delta-function peak in
the interlayer conductance) has yet been observed in experiments.
This suggests two possibilities: (i) The true ground state of the
system at $T=0$ is ferromagnetically ordered, but current experiments
have probed only the $T>T_{KT}$ regime. (ii) The true ground state is
quantum disordered due to some combination of quantum
fluctuations/quenched disorder.

Even in the more conventional possibility (i), there are several
natural sources for the dissipation seen in experiments. The
dissipation could be due to unbound vortices in the hydrodynamic
transport regime\cite{damle-subir} $\hbar \omega\ll k_bT$, and/or due
to the low-energy Griffiths $Q$-modes.

A second aspect of disorder is that as one crosses domains of
different topological density, the spin wave velocity varies
sharply. This is expected to lead to chaotic reflections of spin waves
leading to a diffusive-like behavior at macroscopic
length-scales\cite{sajeev-john}.

Yet another aspect concerns the critical counterflow velocity, which
determines the critical counterflow current. In regions where the
$Q$-mode has a very small gap, the critical velocity will also be
small, and nonlinear/dissipative effects\cite{bilayer-th3} will be
visible at very tiny counterflow current. Similar physics holds for
the critical interlayer tunneling current density. Some recent work
takes the point of view that perhaps the experiments are in the
$T<T_{KT}$ regime, but the smallness of the critical tunneling current
density is a primary mechanism driving the observed
dissipation\cite{su-macd}.

Let us now turn to issues of zero-temperature physics. It is possible
fluctuations quantum disorder the ground state (even in the
periodic-potential model) near the mean-field ground state transitions
that we found. Clearly, one needs an effective theory of the low-lying
modes and their interactions to address these issues. An important
source of quantum disordering is the tunneling of (multiple)
vortices. Due to the spin-charge relation, the quantum tunneling
events (called hedgehogs) of a single vortex will violate charge
conservation. However, especially near transitions, one can imagine
multiple vortices in a unit cell with total vorticity zero tunneling
together. This has a connection to ideas of deconfined
criticality\cite{shivaji1,vbs,deconfined-criticality}. In this type of
scenario, first described for quantum dimer\cite{shivaji1} models and
quantum antiferromagnets\cite{vbs}, dimers/spin excitations, are
described as composites of monomers/spinons. Single hedgehog events are
forbidden due to lattice symmetries, but multiple hedgehogs are
allowed. In certain well-defined extentions of quantum
antiferromagnets, the multiple-hedgehog events can be shown to be
irrelevant\cite{me-subir-large-N}, leading to a critical point with deconfined
spinons\cite{deconfined-criticality}.

The important difference is that in the picture of the quantum Hall
bilayer presented here, it is not lattice symmetries but the
spin-charge relation\cite{lee-kane,shivaji-skyrmion} which enforces
the absence of single hedgehog events. Thus, if deconfined criticality
were to occur for the periodic potential model, it would also likely
occur in the model with true disorder.

The quantum Hall bilayer remains a rich system which potentially supports a
host of physical behaviors yet to be explored, particularly in the presence of strong
disorder potentials that are almost certainly a feature of their realization
in semiconductor systems.  We believe immersing the
system in a periodic potential offers a window through which one may begin
an exploration of this physics.

GM wishes to thank Leon Balents for illuminating conversations, and
also the Kavli Institute for Theoretical Physics at Santa Barbara and
the Aspen Center for Physics for their hospitality when some of this
work was carried out. JS and HAF are grateful to the NSF for partial
support under DMR-0704033, while GM and NB-A wish to thank the NSF for
partial support under DMR-0703992.


\begin{thebibliography}{99}
\bibitem{caution-critical} It should be noted that even below the ``critical''
interlayer tunnelling current the $I-V$ curve has a nonzero slope in
the measurements of Ref.\cite{tiemann-area-law}. It is not clear that
the quantity being measured is the analog of a critical current in a
Josephson junction.
\end{thebibliography}
\end{document}